\documentclass{optica-article}

\usepackage{lineno}
\usepackage{graphicx}

\usepackage{upgreek}
\usepackage{xcolor}
\usepackage{hyperref}
\usepackage{cleveref}

\hypersetup{
  colorlinks  = true,
  linkcolor   = blue,
  citecolor   = blue,
  urlcolor    = blue,
  breaklinks  = true,
}

\begin{document}
\pagestyle{plain}
\title{Photon Number Correlations in Harmonics Driven with Classical and Quantum Light}

\author{Katarzyna M. Kowalczyk,\authormark{1,*} Michael T. Weil,\authormark{1,2} Kevin Murzyn,\authormark{1} Andr\'e Staudte,\authormark{1} Jeff S. Lundeen,\authormark{2} David M. Villeneuve,\authormark{1} and Giulio Vampa\authormark{1}}

\address{\authormark{1}Joint Attosecond Science Laboratory (JASLab), National Research Council of Canada and University of Ottawa, 100 Sussex Drive, Ottawa, Ontario K1A 0R6, Canada\\
\authormark{2}Department of Physics, University of Ottawa, Ottawa ON K1N 6N5, Canada}
\email{\authormark{*}katarzyna.kowalczyk@nrc-cnrc.gc.ca}

\begin{abstract*} 
Recent reports of nonclassical intensity correlations in harmonic generation raised questions about whether strong harmonic correlations necessarily imply genuinely quantum behaviour. Here, we compare the low-photon-number perturbative harmonic generation driven by classical and bright squeezed vacuum sources. The latter generates strong auto- and cross-correlations in the harmonic spectrum. Cauchy-Schwarz inequality and heralding measurements show that the cross-correlations are fully classical and can be explained by the common fluctuations from the driving field. A visible scaling of the measured correlation function with mean harmonic flux is observed, which is well-explained by a click detector-response model. 
\end{abstract*}

\section{Introduction}

Strong-field nonlinear light-matter interactions, like high harmonic generation (HHG), are typically treated semi-classically~\cite{Corkum1993,lewenstein1994a}, where the electron and atom responses are described quantum mechanically while the intense driving field is assumed to be fully classical. However, lately, the possible quantum features of HHG have emerged as a direction in research~\cite{Gorlach2020}. The merging of quantum optics and attosecond science could open a door towards applications like extreme ultraviolet quantum light~\cite{Sloan2023, Gorlach2020, Tzur2024}, provide sub-shot-noise sensitivity in attosecond metrology~\cite{Lemieux2025} or realise quantum randomness as an additional degree of freedom for strong-field spectroscopy.

Recent experiments demonstrated that it is possible to drive strong-field processes like harmonic generation~\cite{Spasibko2017, Rasputnyi2024, Lyu2026} or multi-photon electron emission~\cite{Heimerl2024, Heimerl2025a,Jiang2026,Liu2026a} with quantum light, in particular, a bright squeezed vacuum (BSV). The BSV enhances harmonic generation through its large photon-number fluctuations, resulting in higher harmonic yield~\cite{Spasibko2017, Rasputnyi2024} and a predicted higher cutoff energy~\cite{Gorlach2023}. The high photon-number events associated with the sparse distribution of BSV also provide access to free carrier dynamics beyond typical damage thresholds~\cite{Rasputnyi2024}. Using BSV as a perturbation to classically-driven HHG in spectroscopic studies was also shown to generate nonclassical light confined to spectrally distinct sidebands~\cite{Lemieux2025,Mor2026,Tzur2025}.

The realisation of quantum-light-driven HHG also raised the question of inherent quantum features in high harmonic generation driven by a classical pump~\cite{Gorlach2020,Gombkoto2021, Pizzi2023, Theidel2024, Lange2024, Yi2025,Gombkoto2026}. Recent experiments reported that solid harmonics driven by a pump in a coherent optical state (i.e. 'classically-driven') might exhibit multimodal nonclassical auto- and cross-correlations in the low photon-number regime~\cite{Theidel2024, Theidel2025, Theidel2026}. In particular, when pumped by a nominal coherent-state (such as is emitted by a laser), measurements of the second-order correlation function of the harmonics revealed super-bunching within individual orders (auto-correlation or intra-harmonic correlation) and between distinct harmonics (cross-correlation or inter-harmonic correlation)~\cite{Theidel2024}. Cross-correlations between distinct harmonics were found to consistently violate the Cauchy-Schwarz inequality, an indication that the correlations are nonclassical in nature~\cite{Theidel2024}. In contrast, for higher harmonic orders, when conditioning the measurement of one order on the measurement of another spectrally distinct order, the resulting heralded state was shown to exhibit sub-Poissonian photon-number statistics~\cite{Theidel2026}, another sign of nonclassicality. 

While the quantum features have been reported in the low-photon-number regime for classically-driven harmonics, for BSV-driven HHG the super-bunching of harmonics~\cite{Rasputnyi2024} or sidebands~\cite{Lemieux2025} was proposed to originate from stochastic fluctuations of the driver, with no true quantum witness reported thus far. 

To date, BSV experiments have not yet established whether these harmonic correlations~\cite{Lyu2026} originate from the process itself or are inherited from the pump fluctuations.

In this work, we investigate the  BSV-driven perturbative harmonics in solids, in the low-photon-number regime with click detectors, and compare them to a classical driver. We probe the photon-number statistics and characterise the degree of both auto- and cross-correlation of harmonic 3 (H3) and 5 (H5) as a function of approximate harmonic flux. Finally, we study the effect of heralding on the measurement of harmonics, and test for nonclassicality using the Cauchy-Schwarz inequality for both classically- and BSV-driven harmonics. 
As expected~\cite{Spasibko2017, Lyu2026}, BSV-driven harmonics in the low-photon-number regime are found to be super-bunched ($g^{(2)}>2$) across the measurement range. Importantly, Cauchy-Schwarz tests reveal no violation of the classical bound for the measured inter-harmonic correlations, and while the heralded correlations reduce the bunching significantly, the $g^{(2)}$ is not below 1, i.e. harmonics are not correlated below the shot noise. 
However, in contrast with earlier work~\cite{Theidel2024}, the classically-driven harmonics show no measurable bunching and reveal no significant violation of the tested nonclassicality criteria.
Surprisingly, in the BSV-driven case, the correlations exhibit a strong nonlinear reduction for higher harmonic flux. Because the higher-order coherence functions of BSV are independent of its mean photon number, no strong photon-number dependence is expected in perturbative harmonics from the inherited pump statistics alone. We investigate the possible origins of this scaling and find that it is likely due to the use of on/off detector to measure strongly bunched statistics.

\section{Experimental Details}

\begin{figure}
\centering
\includegraphics[width=0.9\linewidth, trim = 1.6cm 22cm 2.1cm 0, clip]{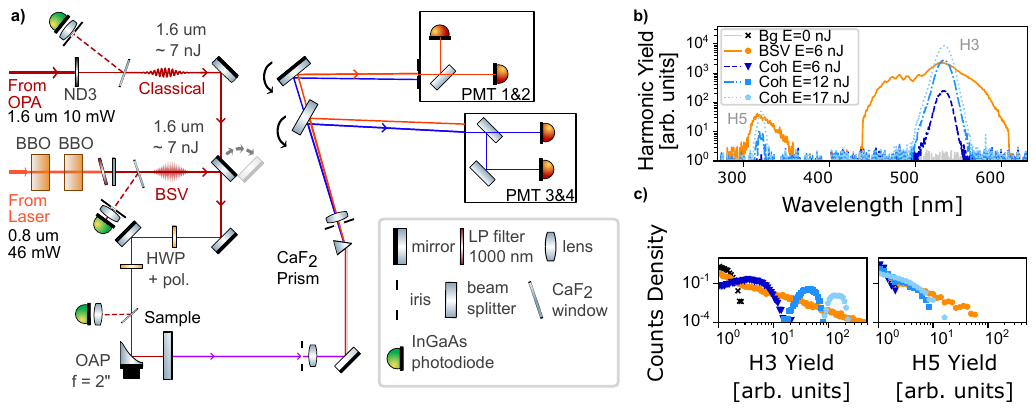}
\caption{Experimental setup and high-photon-number characterization of harmonics. (a) Optical setup for harmonic generation driven by either a classical optical driver or bright squeezed vacuum (BSV). OPA: optical parametric amplifier; BBO: $\beta$-BaB\textsubscript{2}O\textsubscript{4}; LP: long-pass filter; HWP: half-wave plate; OAP: off-axis parabolic mirror; PMT: photomultiplier tube. (b,c) Harmonics generated in Si sample driven by BSV (orange, 6 nJ) or the classical driver (blue, 6, 12, and 17 nJ), measured with an IsoPlane spectrometer. (b) Averaged harmonic spectra. (c) Single-shot wavelength-integrated harmonic-yield distributions for H3 and H5. Black crosses denote background.}
\label{fig:setup}
\end{figure} 

Figure~\ref{fig:setup}a presents our experimental setup. It provides a choice of two drivers for the HHG process: a strong coherent laser beam (classical driver) and a source for BSV. Both sources are produced from a Ti:Sapphire laser system (1~kHz, 800~nm, 50~fs up to 1.8~mJ). The classical driver configuration is the signal output at 1600 nm from an optical parametric amplifier (OPA, Light Conversion TOPAS-Prime) in the degenerate regime, attenuated with a neutral density filter (ND). The BSV source uses a double BBO configuration \cite{Perez2014} separated by 15~cm, pumped with up to 50~$\upmu$J of 800 nm, producing up to 10 nJ of degenerate squeezed vacuum at 1600 nm in a single spatial mode. Either of these drivers is focused ($w_0\approx$20 ~$\upmu$m) into thin-film silicon (500 nm) at the back surface of a sapphire substrate, which avoids propagation effects in the sample. The power of the driver (pump) is controlled with a half-wave plate (HWP) and polariser pair. 

In this work, we distinguish between high- and low-photon-number regimes. In the high-photon-number regime, when the harmonic signal is measured as a continuous intensity, the harmonic yield serves as a proxy for a photon number. In the low-photon-number regime, where individual detection events are recorded, the photon statistics are extracted from multiplexed single-photon-counting measurements. 

Figure~\ref{fig:setup}b compares the spectrum of H3 and H5 at each driver, measured after the sample in the high-photon regime. The two drivers are neither matched in the bandwidth ($\Delta \omega_{\text{BSV}} > 4 \Delta \omega_{\text{classical}}$) nor in energy, hence the difference in harmonic bandwidth. For the correlation experiments, we match the mean flux of the harmonics and not the flux of the pump. Single-shot measurements of the harmonic flux with a commercial spectrometer (Isoplane 320 with diffraction grating and Pi-Max4 ICCD camera) provide access to the intensity fluctuations of the harmonics in the high-photon-number regime, shown in Fig.~\ref{fig:setup}c. The harmonics driven by BSV (orange markers) exhibit a broader intensity distribution than the classically-driven harmonics (blue markers), a sign of a bunched state.

To access the harmonic low-photon-number statistics directly, we use four separate PMTs arranged in a double Hanbury-Brown-Twiss (HBT) configuration. With four PMTs, we can measure correlations $g_{ij}^{(n)}$ between harmonics $i$ and $j$ up to $n=2$. The harmonic beam is first sent onto a CaF\textsubscript{2} prism, spatially separating the harmonic wavelengths, and then re-imaged onto an iris after the first beam splitter. The tilt of the mirrors before the PMTs controls the wavelength sent onto each PMT pair. Because of the spatio-spectral separation out of the prism, the small aperture before the PMT pairs also acts as a spectral filter, allowing only a bandwidth of a single harmonic to pass through. The harmonic $i$ flux is monitored by the mean click number $\langle n_i \rangle$, defined as the average number of total clicks in the PMT pair and is used as a proxy for the mean photon number in each harmonic, albeit not efficiency-corrected. For further details on the photon-counting data analysis, see Supplemental~\ref{SM:anal}.

\section{Nature of inter-harmonic correlations}
\begin{figure}
\centering\includegraphics[width=0.7\textwidth, trim = 0cm 4cm 0cm 0, clip]{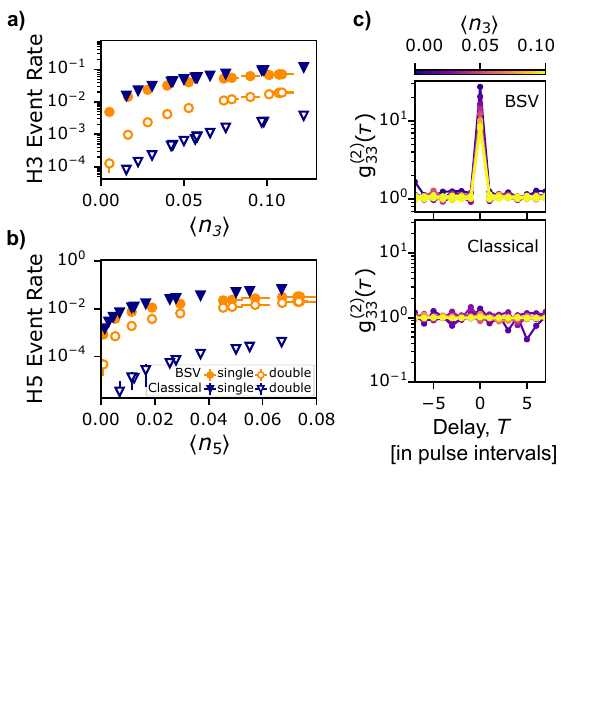}
\caption{Event rate per laser shot for H3 (a) and H5 (b) as a function of mean click number $\langle n_i\rangle$. Filled markers: exactly one PMT click; open markers: double-click events. BSV-driven harmonics (orange) exhibit substantially higher double-click probabilities than classically driven harmonics (blue). Error bars are present for all points but are generally smaller than the markers. (c) Second-order correlation function $g^{(2)}_{33}(T)$ for H3 driven by BSV (top) and classical light (bottom). Colours correspond to mean click number.}

\label{fig:event}
\end{figure}

Looking first at event rates, we gain initial insight into the harmonic correlations present for different drivers. Figure~\ref{fig:event}a and b show the rate of single (filled markers) and double (empty markers) events as a function of harmonic flux for H3 and H5, driven by either BSV (orange markers) or the classical pump (blue markers). While at the same flux, the single-event rate between BSV and classical drivers is approximately matched, the double-event rate is orders of magnitude higher for the BSV harmonics, i.e. the probability of photons arriving in pairs is higher in BSV-driven harmonics than in the classical case. The second-order coherence (or correlation) function $g^{(2)}$ more formally quantifies the probability of coincidences. Here, each of the PMT pairs performs an intensity correlation measurement. Because of the pulsed nature of the source, we define the time-dependent $g^{(2)}(T)$ as a function of the delay of the click sequence $n_{j}$ by m pump-pulse intervals relative to $n_{i}$ ($T = mT_{\text{rep}}$ for $m = 0,1,2 \ldots$ and $T_{\text{rep}}=$ 1 ms):
\begin{equation}
\label{eq:g2}
    g_{ij}^{(2)}(T) = \frac{\langle n_i(0) n_j(T) \rangle}{\langle n_i(0)\rangle\langle n_j(T)\rangle},
\end{equation}
where the expectation values $\left\langle \right\rangle$ are over the recorded sequence. Figure~\ref{fig:event}c shows the measured $g^{(2)}(T)$ for H3 driven by a BSV (top) and a classical (bottom) source as defined in Equation~\ref{eq:g2}. The colours of the lines are mapped to the mean click number $\langle n_3\rangle$. In contrast with previous work~\cite{Theidel2024}, the measured second-order correlations of classically-driven harmonics (blue triangles) show no statistically significant deviation from unity. Although this is consistent with a Poissonian distribution, it does not necessarily imply a classical state \cite{Gerry2004}. However, for the BSV-driven harmonics, $g^{(2)}(T=0)$ deviates strongly from 1, increasing by an order of magnitude and indicating strong bunching. An equivalent behaviour is seen in H5 (see Supplementary Figure~\ref{figSM:gT}). This shows that BSV-driven harmonics exhibit large photon-number fluctuations absent in the classically-driven case, and larger than those of the BSV driver ($g^{(2)}_{\text{BSV}}\approx2.3$). 

\begin{figure}[t]
\centering\includegraphics[width=0.6\linewidth, trim = 0cm 1.5cm 0cm 0,clip]{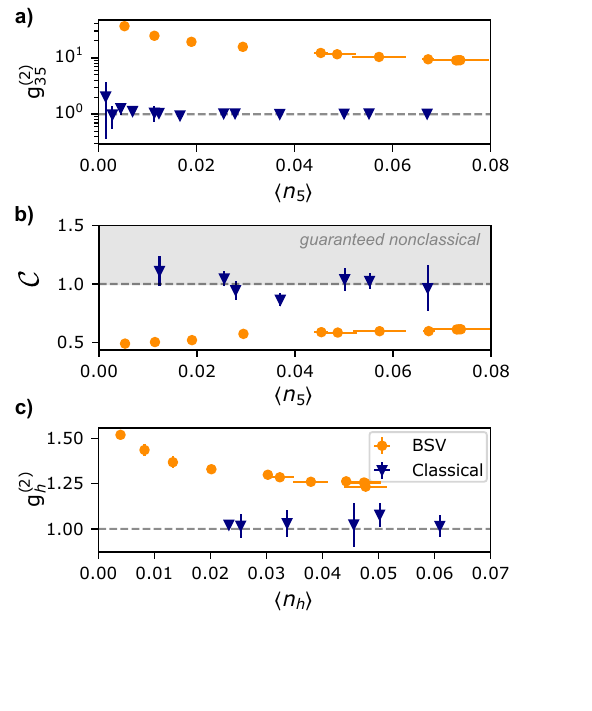}
\caption{Cross-correlation measurements of H3 and H5. (a) Inter-harmonic correlation $g^{(2)}_{35}$ as a function of mean click number. (b) Cauchy-Schwarz ratio; values above unity violate the classical bound. (c) Heralded correlation of H3 conditioned on detection of H5 as a function of the mean click number in the heralded arm. Error bars are present for all points but are often smaller than the markers.  All three panels are derived from the same dataset, acquired from the same measurement, collection, and attenuation settings, and differ only in their analysis.}
\label{fig:cross}
\end{figure}

The four PMT beamsplitter tree setup allows for the measurement of inter-harmonic coherence, defined by the cross-correlation $g^{(2)}_{35}(T)$. As the only divergence from unity occurs at $T=0$, we will focus on $g^{(2)}(T=0)$ ($\equiv g^{(2)}$). Figure~\ref{fig:cross}a presents the cross-correlation between H3 and H5, i.e. $g^{(2)}_{35}$. For the classically-driven harmonics, it remains 1 within experimental uncertainty, while for the BSV-driven harmonics, a strong correlation ($g^{(2)}_{35}\gg1$) appears across the explored harmonic flux range. The large value of inter-harmonic $g^{(2)}$ points towards strong correlations in the intensity fluctuations of the two modes, albeit, the nature of the correlations is not necessarily nonclassical. 

The Cauchy-Schwarz inequality, is often used as a probe of nonclassical correlation origin~\cite{Wasak2014, Gombkoto2021, Theidel2024}. We use the Cauchy-Schwarz ratio $\mathcal{C}$ for two beams $i$ and $j$~\cite{Loudon2000, Gerry2004}:
\begin{equation*}
    \mathcal{C} = \frac{(g_{ij}^{(2)})^2}{g_{ii}^{(2)}g_{jj}^{(2)}}.
\end{equation*}
The $\mathcal{C}$ ratio gives an upper bound on the classical correlation: $\mathcal{C}\leq 1$. While the classically-driven harmonics in Figure~\ref{fig:cross}b sit exactly at that boundary, the BSV-driven $\mathcal{C}$ is below 1, and so the fluctuation correlations are fully compatible with a classical description. 

Heralded coincidence counting provides an additional test of nonclassical photon statistics~\cite{Pearce2026, Tsao2025}. In systems that generate strongly correlated photon pairs (e.g. two-mode squeezed vacuum), if we condition (herald) the presence of one mode on the detection of another, the heralded measurement can produce sub-Poissonian statistics ($g^{(2)}$$\leq 1$), a sign of nonclassicality~\cite{Wang2021b}. We can similarly investigate if conditioning H3 ($n_{i,j}$) on the presence of H5 ($n_h$) reveals signatures of nonclassicality:

\begin{equation*}
    g_h^{(2)}= \frac{\langle n_hn_in_j \rangle\langle n_h \rangle}{ \langle n_in_h\rangle\langle n_jn_h\rangle}.
\end{equation*}
Figure~\ref{fig:cross}c shows the result of heralding the $g^{(2)}$ of H3 on the presence of H5. We present the results for both BSV- and classically-driven harmonics for completeness, even though the latter shows negligible cross-correlation in Figure~\ref{fig:cross}a. 
The heralded $g^{(2)}$ approximates the classical case ($g^{(2)}_{h}  \approx 1$), but never decreases below 1, which is consistent with the classical nature of the inter-harmonic correlations. Thus, the fluctuations in one harmonic are almost entirely suppressed by conditioning it on the presence of the other, but are not less than the shot noise. Ultimately, while BSV-driven harmonics generate large harmonic cross-correlations, the measurements provide no evidence that a nonclassical description of these correlations is necessary. 

\begin{figure}
\centering
\includegraphics[width=0.5\linewidth, trim = 0cm 4.3cm 0cm 0, clip]{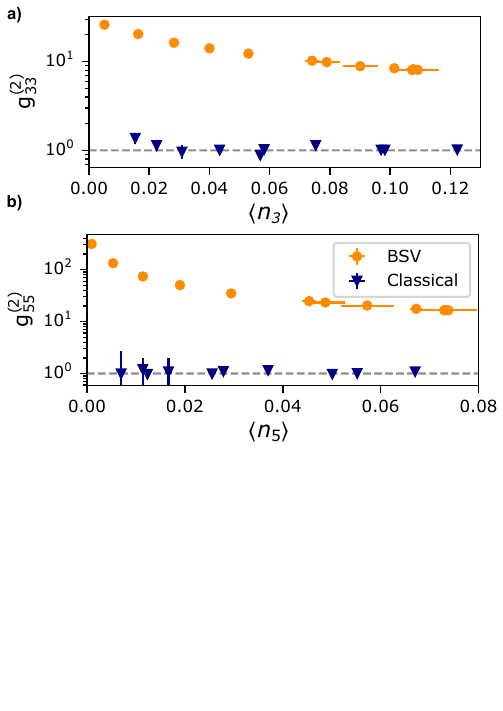}
\caption{Autocorrelation measurements of Si harmonics driven by BSV and a classical optical driver. (a,b) $g^{(2)}_{35}$ as a function of mean click number for H3 and H5, respectively. Horizontal and vertical error bars are present for all points but are often smaller than the marker size.}
\label{fig:auto}
\end{figure}

Previous reports of super-bunching of harmonics driven by BSV~\cite{Spasibko2017, Rasputnyi2024} have not investigated the nonclassicality of the fluctuations directly but have pointed towards the amplified fluctuations of the pump as the origin of the heavy-tailed photon number distribution of the harmonics. High-order sidebands of a coherent high-harmonic spectrum perturbed with a BSV were also reported to have a $g^{(2)}$~\cite{Lemieux2025} comparable to the BSV's $g^{(2)}$, highlighting the direct mapping of the fluctuations in the linear perturbation. While the differing values of the harmonic statistics have been attributed to different generation mechanisms~\cite{Lyu2026}, our results show that strong inter-harmonic correlations and high $g^{(2)}$ values do not always point towards quantum features.

Finally, the super-bunching in the low-photon-number regime was observed in BSV-driven harmonics, but the classical driver correlations remain at the classical baseline of $g^{(2)}=1$, contrary to previous work~\cite{Theidel2024}. With a comparable driving wavelength and at equivalent intensities (Keldysh parameter: $\gamma > 5$ for all points), we do not find any increase in harmonic correlations with decreased driver intensity or any nonclassical features in the Cauchy-Schwarz or heralded measurements. 
While this could be due to a number of other differences, our measurement shows that the previously reported nonclassical features appear sensitive to experimental conditions and might not be a universal feature of low-photon-number solid-state HHG. 

\section{Photon-number scaling}
Figure~\ref{fig:cross} establishes that the observed inter-harmonic correlations are compatible with a classical description.
However, an additional feature of the data remains unexplained: the measured inter-harmonic correlation function evolves with harmonic flux, i.e. mean click number. This scaling is also visible in the intra-harmonic measurements: Figure~\ref{fig:auto}a and b present the autocorrelation function of H3 and H5, respectively, as a function of the mean click number. The classically-driven harmonics' second-order correlations remain 1 for the full range of $\langle n_i\rangle$, but the $g^{(2)}$ of the BSV-driven harmonics increases nonlinearly with decreasing harmonic flux. The $g^{(2)}$ of H3 is consistent with a square-root scaling $g^{(2)}_{33} \propto \langle n_3\rangle^{-0.54}$ ($R^2$ = 0.990), while the H5 displays stronger suppression of bunching: $g^{(2)}_{55} \propto \langle n_5\rangle^{-0.83}$ ($R^2$=0.996), see Extended Data Figure~$S3$. 
The scaling appears in both auto- and cross-correlation, thereby suggesting a robust mechanism behind the trend. 

The second-order correlation function $g_{n\omega}^{(2)}$ of a perturbative n\textsuperscript{th} harmonic can be expressed in terms of the higher-order coherence functions of the driver~\cite{Teich1966,Spasibko2017,Rasputnyi2024}: 
\begin{equation}
\label{eq:gn}
g_{n\omega}^{(2)} = \frac{g_{\omega}^{(2n)}}{(g_{\omega}^{(n)})^2}.
\end{equation}
where $g_{\omega}^{(n)}$ is the n\textsuperscript{th}-order coherence function of the pump. For BSV (in the large-photon-number limit), the n\textsuperscript{th}-order coherence is large ($g^{(n)} = (2n-1)!!\gg1$) \cite{Qu1995, Allevi2015, Spasibko2017, Rasputnyi2024} and independent of the average number of photons in the BSV. It follows that if the only reason for the super-bunched nature of harmonics was the photon statistics of the source, the correlation function of the harmonics would not evolve with the average photon number, contrary to our measurements. 

The robustness of the decrease across intra- and inter-harmonic correlations and its persistence in the heralded measurement suggest an origin beyond the common fluctuations. One possible contribution arises from the click nature of the PMTs. The heavy-tailed distribution of BSV-driven harmonics increases the possibility of a rare multi-photon event that cannot be resolved by two click detectors, leading to a possible reduction in the measured correlations with on/off detectors. 
\begin{figure}[t]
\centering\includegraphics[width=0.5\linewidth, trim = 0cm 7.5cm 2.5cm 0, clip]{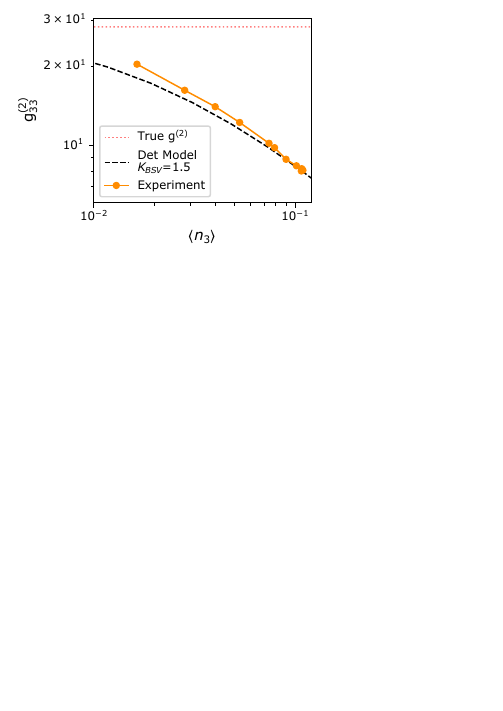}
\caption{Modelling of the multiplexed on/off detector's effect on the second-order correlation function $g_{33}^{(2)}$ of BSV-driven H3 (dashed black line) compared with experimental data (orange circles). The expected value of $g^{(2)}$ is extracted from the higher-order moments of the modelled BSV driver distribution: $g^{(2)}_{3\omega} = g^{(6)}/(g^{(3)})^2=28.4$ according to Equation~\ref{eq:gn}.}
\label{fig:model}
\end{figure}

We computationally investigate this click detector effect on the measured correlations of a super-Poissonian state of light. Assuming both detectors have equal efficiency and a 50:50 splitting, we calculate the probabilities of empty, single and double events for any incoming photon number. Then, we model the total detector response by scaling these event probabilities by a photon-number distribution of a perturbative third harmonic generated by a BSV with $g^{(2)}=2.3$ (see Supplementary ~\ref{SM:model}). 
The modelled $g_{33}^{(2)}$ scaling is shown in Figure~\ref{fig:model} (black dashed line), along with the measured values (orange markers). The model shows good agreement with experimental data. While some residual differences remain, the reproduced three-fold reduction of $g^{(2)}$ suggests the detector response can largely explain the visible scaling. While the current model does not exclude additional contributions to the scaling, these findings highlight that measurement with multiplexed on/off detectors of highly nonlinear processes, which result in super-bunched light, can present with erroneous modulation of the statistics. This is a significant effect even at very low mean count rate (triple click probability $P_3\ll 0.1\%)$ and while the underlying light source statistics remain fixed. Therefore, detector response must be taken into careful consideration when interpreting photon-number-dependent correlations of highly bunched fields. 

\section{Conclusions}

We presented a comparative study of intensity fluctuations in solid harmonic generation driven by both classical and BSV light in the low-emitted-photons regime. While the classically-driven harmonics have second-order correlations consistent with unity, the BSV-driven case results in large harmonic fluctuations, and we measure super-bunching. The measured inter-harmonic correlations do not violate the nonclassicality criteria tested here.
An observed photon-number scaling of the correlation function can be closely replicated by the detector effect, which highlights the limitation of the use of on/off detectors to characterise strongly superbunched light from highly nonlinear processes.

\section{Back matter}
\begin{backmatter}
\bmsection{Funding}

\bmsection{Acknowledgment} 
The authors would like to acknowledge David Crane and Ryan Kroeker for continuing technical support, as well as Dr. Kyle Jordan, Dr. Philip Bustard and Dr. Frédéric Bouchard for productive scientific discussions. G.V. acknowledges funding from NSERC Discovery grant (RGPIN-2021-04286). J.L. acknowledges funding from NSERC Discovery grant (RGPIN-2026-06045).

\bmsection{Disclosures} The authors declare no conflicts of interest.

\bmsection{Data availability} Data underlying the results presented in this paper are not publicly available at this time but may be obtained from the authors upon reasonable request.

\bmsection{Supplemental documents}
See the Supplemental Document for additional details on data analysis, extended data and the detailed explanation of the model.

\end{backmatter}

\bibliography{QuantumOptics}

\appendix

\section{Details of the data analysis}
\label{SM:anal}

To investigate the low-photon-number regime we need photon counting detectors. Here, we use PMTs (Hamamatsu H9305-04) in photon-counting mode~\cite{Jonsson2019, Ota2021} for their broad spectral response, with sensitivity down to 200 nm~\cite{Hamamatsu2017PMT}. We run them in the high-gain photon-counting regime. Each PMT signal is sent into $50\,\Omega$ pre-amplifier and collected with a digitiser (Spectrum Instrumentation M4i.2234-x8) triggered by the laser. The signal is then integrated over 20 $\upmu s$  to account for delay box jitter and the temporal shape of the electronic PMT signal (1 ms is the time between consecutive laser shots). To differentiate between the presence and absence of a photon, we measure a dark background that defines the distribution of no-photon signal and extract a threshold above which we assume a click for every PMT. 

The data extracted contain the binary value of 0 for no photon and 1 for photon present. The sequence of clicks for a specific intensity and all 4 four detectors is then analysed, calculating the number of shots, the average photon counts for each PMT, the second-order correlations between PMTs, and the third-order correlation. To estimate the statistical uncertainties, the data within each intensity group are divided into five smaller subsets. All the quantities are calculated independently for each subset, and the mean and standard deviation across the subsets are then used to obtain the plotted values and their corresponding error bars.

\section{Extended Data}
\label{SM:data}

Here, we provide additional measurement data to further validate the claims of the paper.

\begin{figure}[h]
\centering\includegraphics[width=0.8\linewidth, trim = 0cm 5cm 0cm 0, clip]{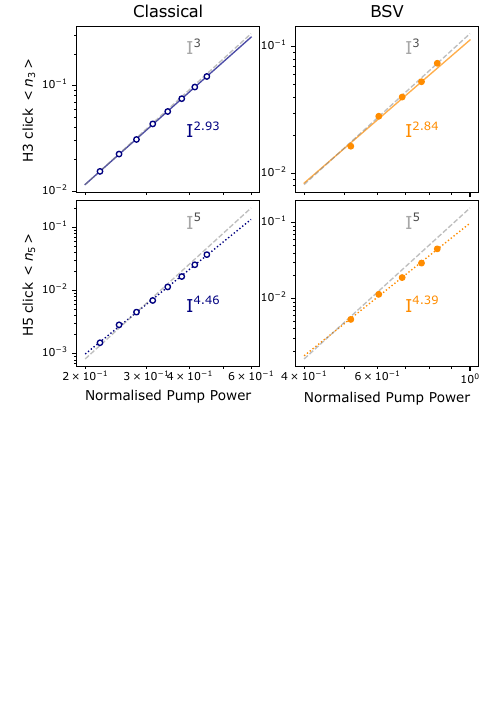}
\caption{Harmonic intensity scaling with the incoming pump power. Top panels show the scaling for H3 and bottom for H5 for the classically-driven (left) and BSV-driven (right) regimes. The scaling parameters, for H3: $2.93\pm 0.01$ (classical) and $2.84\pm0.12$ (BSV) and H5: $4.46\pm0.05$ (classical) and $4.39\pm0.11$ (BSV) are consistent between the driving regimes and in reasonable agreement with perturbative harmonic picture.}
\label{figSM:Iscaling}
\end{figure}

Figure \ref{figSM:Iscaling} presents the mean click scaling of the measured harmonics with the pump power. The experimentally observed scaling exponents (2.9 and 4.5 for the third and fifth harmonics, respectively) are in reasonable agreement with the perturbative predictions (3 and 5). Despite the small deviation, the overall scaling behaviour and the Keldysh parameter $\gamma \gg 1$ for all measurements are consistent with harmonic generation in the perturbative regime.

\begin{figure}[h]
\centering\includegraphics[width=0.8\linewidth, trim = 0cm 5.5cm 0cm 0, clip]{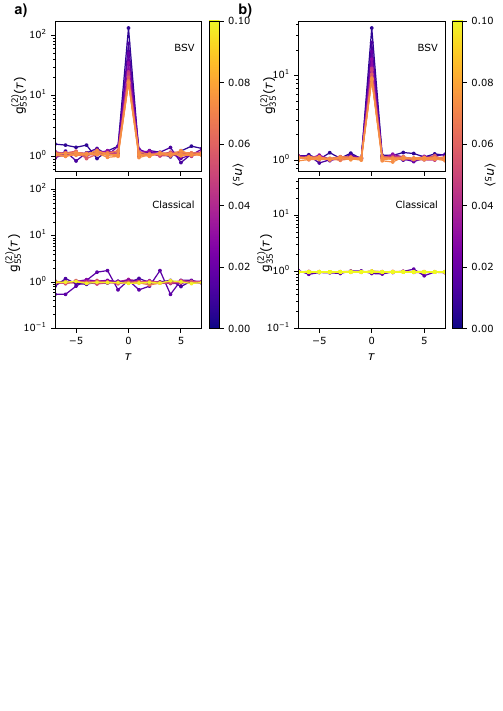}
\caption{The second-order coherence function of a) harmonic 5 ($g^{(2)}_{55}(T)$) and b) inter-harmonic 3 and 5 ($g^{(2)}_{35}(T)$), driven by BSV (top) and classical light (bottom) as a function of the shift between the laser shots $T$. Different line colours represent the mean click number in the measured harmonic (see colour bar).}
\label{figSM:gT}
\end{figure}

Figure~\ref{figSM:gT} shows analogous bunching at $T=0$ as in Figure~\ref{fig:event}c) for H5 and inter-harmonic measurement. As with H3, the second-order correlation function only deviates from unity at $T=0$. The exception is the classically-driven $g^{(2)}_{55}$ (top left) for the lowest mean click number $\langle n_5 \rangle \approx 0.01$, which fluctuates around 1. We attribute this behaviour to the relatively low total number of coincidence counts, resulting in large uncertainty.  

\begin{figure}[h]
\centering\includegraphics[width=0.8\linewidth, trim = 0cm 6.5cm 0cm 1.5cm, clip]{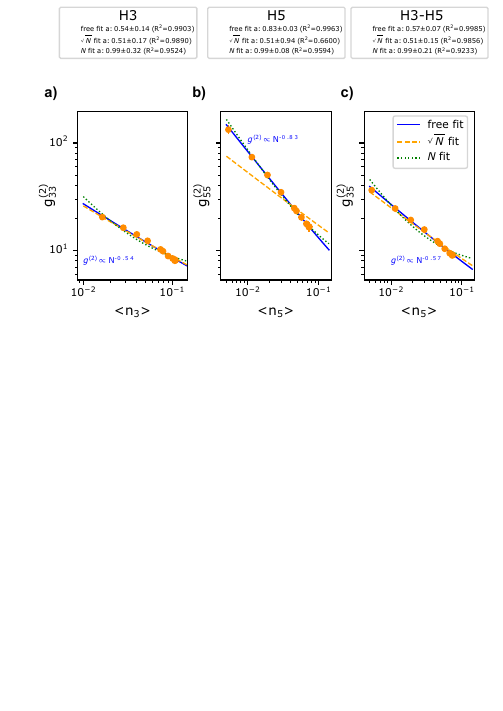}
\caption{BSV-driven $g^{(2)}$ mean click-number scaling (orange circle) for a) harmonic 3, b) harmonic 5 and c) inter-harmonic 3 and 5. The extracted free parameters (blue solid) are $0.54\pm0.14$, $0.83\pm0.03$ and $0.57\pm0.07$ respectively. Fits with the specified scaling parameters of $1/\sqrt{N}$ (orange dashed line) and $1/N$ (green dotted line) for comparison.}
\label{figSM:fit}
\end{figure}

The fitted mean click-number scaling parameter for each second-order correlation function $g^{(2)}(N) = 1 + B/N^{C}$ and the extracted nonlinear scaling parameter $C$ are presented in Figure~\ref{figSM:fit}. Due to the low dynamic range of the measurement (only one order of magnitude in click-number), we present the fit alongside $N^{-1}$ and $N^{-1/2}$ for completeness. 

Table~\ref{tab:R2} compares how well the above fits can reproduce the data with a coefficient of determination $R^2$.
\begin{table}[]
    \centering
    \begin{tabular}{c|c|c|c|}
          Fit &  $g^{(2)}_{33}$ &  $g^{(2)}_{55}$ &  $g^{(2)}_{35}$ \\ 
          
         \hline
         Free fit & 0.990 & 0.996 & 0.999 \\
         $\sqrt{N}$ fit & 0.989 & 0.660 & 0.986 \\
         $N$ fit & 0.952 & 0.959 & 0.923 
    \end{tabular}
    \caption{Coefficient of determination $R^2$ for second-order correlation scaling with the mean click number }
    \label{tab:R2}
\end{table}

\section{Photon-number scaling model}
\label{SM:model}
To investigate the origin of the scaling, we model the response of an on/off detector tree to the generated harmonic distribution. The model consists of two steps. Firstly, we calculate the photon-number distribution of the generated harmonic by averaging over the pump intensity distribution. Secondly, we calculate the response of the click detectors to this harmonic photon-number distribution.  

To extract a probability of the third harmonic, we model a high photon number BSV distribution using a continuous $\Gamma$ distribution (from the \texttt{SciPy.stats} module in Python):
\begin{equation*}
f(I,\alpha,\beta) = \frac{\beta^{\alpha}I^{\alpha-1}e^{-\beta I}}{\Gamma({\alpha})}.
\end{equation*}

Here $I$ is the intensity variable, the shape is defined in relation to Schmidt number of the source by $\alpha = K_{\text{BSV}}/2$, and the mean intensity $\mu_I$ defines the scale parameter as $\beta^{-1} = \mu_I/\alpha$. 

We use this approximate $\Gamma$ distribution instead of the exact BSV distribution~\cite{Manceau2019}, because we need to reproduce both the probability of empty shots and the long tail of the large photon-number regime to describe our experimental data. This simplified model uses the $\Gamma$ distribution as an approximation to the statistical distribution of the pump; it is not a description of the full quantum state of the BSV. The continuous distribution is evaluated numerically over a finite intensity range, with the upper intensity limit chosen using a cutoff intensity which probability is below $N_{\text{max}}^{-1}$, where $N_{\text{max}}$ is the maximum number of laser shots collected for any data point in this study. The resulting intensity distribution is normalised and discretised on a numerical grid for the subsequent integration. 

For the perturbative harmonic, the harmonic yield scales as $H=\epsilon I^q$ where q is the order of perturbation and $\epsilon$ is the efficiency (here we used $\epsilon=10^{-6}$). Since both intensity and efficiency have arbitrary units in this model, they can be scaled together to match the measured data. 

To account for the statistical fluctuations in the harmonic generation process, we approximate the probability of generating exactly $n$ harmonic photons for a given pump intensity $I$ using a Poisson distribution:
\begin{equation*}
    P(N=n|I) = \frac{\big(H(I)\big)^ne^{-H(I)}}{n!}
\end{equation*} 

This gives a different conditional photon-number distribution for each pump intensity. The final harmonic photon-number distribution is then obtained by averaging these conditional distributions over the pump intensity distribution $P(N=n|I) = \int_0^{I_{max}} P(n=n|I)f(I)dI$.

In practice, this integration is performed numerically over the discretised intensity grid. The resulting photon-number distribution is normalised over the calculated photon-number range. The parameters for intensity grid spacing, maximum intensity and the maximum number of harmonic photons are all verified for convergence of the model. The model does not take into account different modes within the pump, but it simply assumes that the pump can be fully represented by a single intensity. 

To include detector compression of the harmonic distribution, we model a two-detector tree with equal splitting and detection efficiency of $\eta=$30\%. The detection efficiency only approximates the quantum efficiency of the detector and not the general collection losses. Since the detectors are on/off detectors, they only register whether at least one photon has been detected, independently of the number of photons arriving at the same detector.

For $N$ incoming harmonic photons,  the probability of empty, single and double clicks for $N$ incoming photons is calculated as:
\begin{align*}
    P_0 &= (1-\eta)^N \\
    P_1 &= 2((1-0.5\eta)^N -(1-\eta)^N)\\
    P_2 & = 1 - 2(1-0.5\eta)^N + (1-\eta)^N
\end{align*}

To obtain the detector response for the full harmonic distribution, we average these probabilities over the photon-number distribution $\sum_{N'=0}^{N_{\max}} P(N=N')P_j(N')$ for $j=0,1,2$ corresponding to no click, exactly one click and simultaneous double click, respectively.

We can calculate these click probabilities, the mean click value in each detector $\langle n_{A, B}\rangle=\frac{1}{2}P_1+P_2$, the mean click number $\langle n_q \rangle = \langle n_A\rangle + \langle n_B \rangle$ and $g^{(2)} = \frac{P_2}{\langle n_A\rangle\langle n_B \rangle}$ for any starting distribution. 

This model captures the detector compression effect. The response is nonlinear with respect to the incoming photon number: once a detector has registered a click, additional photons arriving at the same detector do not increase its recorded value. This modifies the second-order correlation function compared with the correlation calculated directly from the incoming photon-number distribution. 

The detector response can recreate the three-fold measured reduction in $g^{(2)}$ seen in Figure~\ref{fig:model}. The detector model matches the results reasonably well even with a simplified detector and harmonic picture, suggesting the experimental limitations themselves could reproduce the measured scaling. 

\end{document}